\documentclass[10.5pt,compsoc]{JSC}
\graphicspath{{figures/}}
\usepackage{graphicx}
\usepackage{footmisc}
\usepackage{subfigure}
\usepackage{url}
\usepackage{multirow}
\usepackage[noadjust]{cite}
\usepackage{amsmath,amsthm}
\usepackage{amssymb,amsfonts}
\usepackage{booktabs}
\usepackage{color}
\usepackage{ccaption}
\usepackage{booktabs}
\usepackage{float}
\usepackage{fancyhdr}
\usepackage{caption}
\usepackage{xcolor,stfloats}
\usepackage{comment}
\usepackage{cuted}  
\usepackage{captionhack}
\usepackage{epstopdf}
\usepackage{times}
\headevenname{}%
\headoddname{{}}%

\newtheoremstyle{mystyle}{0pt}{0pt}{\normalfont}{1em}{\bf}{}{1em}{}
\theoremstyle{mystyle}
\renewcommand\figurename{Fig.~}

\newcommand{\nop}[1]{}

\makeatletter
\renewcommand{\@biblabel}[1]{[#1]\hfill}
\makeatother

\begin{document}

\thispagestyle{empty}



\hyphenpenalty=50000

\makeatletter
\newcommand\mysmall{\@setfontsize\mysmall{7}{9.5}}

\newenvironment{tablehere}
  {\def\@captype{table}}
  {}
\newenvironment{figurehere}
  {\def\@captype{figure}}
  {}

\thispagestyle{plain}%
\thispagestyle{empty}%

\let\temp\footnote
\renewcommand \footnote[1]{\temp{\zihao{-5}#1}}
{}

\begin{strip}
{\center
{\zihao{3}\textbf{
What Contributes to a Crowdfunding Campaign’s Success? Evidence and Analyses from GoFundMe Data}}
\vskip 9mm}

{\center {\sf \zihao{5}
Xupin Zhang, Hanjia Lyu, and Jiebo Luo$^*$, \textit{Fellow, IEEE}
}
\vskip 5mm}
%

\centering{
\begin{tabular}{p{160mm}}

{\zihao{-5}
\linespread{1.6667} %
\noindent
\bf{Abstract:} {\sf
Researchers have attempted to measure the success of crowdfunding campaigns using a variety of determinants, such as the descriptions of the crowdfunding campaigns, the amount of funding goals, and crowdfunding project characteristics. Although many successful determinants have been reported in the literature, it remains unclear whether the cover photo and the text in the title and description could be combined in a fusion classifier to better predict the crowdfunding campaign’s success. In this work, we focus on the performance of the crowdfunding campaigns on GoFundMe across a wide variety of funding categories. We analyze the attributes available at the launch of the campaign and identify attributes that are important for each category of the campaigns. Furthermore, we develop a fusion classifier based on the random forest that significantly improves the prediction result, thus suggesting effective ways to make a campaign successful.}
\vskip 4mm
\noindent
{\bf Key words:} {\sf crowdfunding; fusion; classification; GoFundMe}}

\end{tabular}
}
\vskip 6mm

\vskip -3mm
\zihao{6}\end{strip}

\thispagestyle{plain}%
\thispagestyle{empty}%
\makeatother
\pagestyle{tstheadings}

\begin{figure}[b]
\vskip -6mm
\begin{tabular}{p{44mm}}
\toprule\\
\end{tabular}
\vskip -4.5mm
\noindent
\setlength{\tabcolsep}{1pt}
\begin{tabular}{p{1.5mm}p{79.5mm}}
$\bullet$& Xupin Zhang is with the Warner School of Education and Human Development, University of Rochester, Rochester, NY 14627, USA. E-mail: xzhang72@u.rochester.edu.\\
$\bullet$& Hanjia Lyu is with the Goergen Institute for Data Science, University of Rochester, Rochester, NY 14627, USA. E-mail: hlyu5@ur.rochester.edu.
 \\
$\bullet$& Jiebo Luo is with the Department of Computer Science, University of Rochester, Rochester, NY 14627, USA. E-mail: jluo@cs.rochester.edu.
 \\
$\sf{*}$&
To whom correspondence should be addressed. \\
          &          

\end{tabular}
\end{figure}\zihao{5}

\section{Introduction}
\label{s:introduction}
\noindent
In recent years, the rise of charitable crowdfunding plat- forms such as GoFundMe makes it possible for Internet users to offer direct help to those who need emergency financial assistance. However, the success rate of the campaigns is found to be less than 50\% (success is defined as a campaign that reaches its funding goal) [1]. 

Numerous studies suggest image and text features may have an impact on crowdfunding success. For instance, Yuan, Lau, and Xu [23] developed the Domain-Constraint Latent Dirichlet Allocation (DC-LDA) topic model to effectively extract topical features from texts of the crowdfunding campaigns. In addition, Cheng et al. [24] concluded that the image features could improve success prediction performance significantly. 

The facial expression of emotion can influence interpersonal trait inferences [25]. For instance, the existing literature [26, 27, 28] suggest a positive impact of the smile on interpersonal judgments. Smiling people are, in general, perceived as more sociable, more honest, more pleasant, politer, and kinder. Smile intensity has been documented to influence one’s perceived likability—facilitative effect of smile intensity on warmth [29]. We find that the joint analysis of the textual and visual information has not been fully explored yet in previous studies.

In this study, we analyze GoFundMe, which is currently the biggest crowdfunding platform. This site allows people to raise money for various events, from life events like a wedding to challenging situations such as accidents or illness. We analyze the pages of all 10,974 available crowdfunding campaigns on the site as of November 19, 2019. This research investigates the success determinants for a crowdfunding campaign. Our research questions are: can we quantify the economic returns of the image and text features? If so, can we reliably predict the fundraising’s performance using the attributes available at the launch of the crowdfunding campaign? We investigate how much the difference between successful and unsuccessful campaigns can be explained by text and image factors. We predict crowdfunding outcomes by combining both textual and pictorial descriptions of the crowdfunding projects. This combination provides a more comprehensive view of the factors in successful crowdfunding projects and helps to better take into account possible interrelations.

Using the variables extracted from our dataset, we define the measure of the crowdfunding’s success as the ratio of the current amount of money that has been raised to the fundraiser’s goal amount. To further understand the determinants of successful campaigns, we separately analyze the image and text features to understand how much each of these factors contributes to the success of a crowdfunding campaign.

The main contributions of our research are:
\begin{itemize}
    \item We analyze image and text features that are important to specific categories of crowdfunding campaigns.
    \item We analyze facial attributes in the cover image and examine their impact on crowdfunding performance.
    \item We design a fusion classifier that can combine both textual and pictorial descriptions of crowdfunding projects to reliably predict crowdfunding outcomes.
\end{itemize}

Taken together, this study is among the first to adopt both text features and image features from project descriptions and cover images to analyze and predict the GoFundMe crowdfunding projects’ success. We focus on GoFundMe because campaigns launched here are charity-minded and rich in categories, while campaigns launched on Kickstarter are exclusively entrepreneurial. This project provides a more comprehensive view of the factors in successful project funding of projects and better take into account possible interrelations. The managerial implication of our research is that crowdfunding platforms can better identify the most influential image and text features. They can offer strategic suggestions to help their users (fundraisers) raise more money quickly and also attract more donors to their websites.

\section{Related Work}
\label{s:related_work}
\noindent
Many studies have been conducted to explore the determinants of campaign success on the crowdfunding platform Kickstarter. It has been found that active communications with the platform members [2], project description and image [3], individual social capital as proxied by the number of contacts on social networks [4], geographical distance [5], linguistic style [6], the amount of the funding goal [7], project duration  [8], and the content of the project updates [9, 10] have a significant impact on the success of crowdfunding projects.

Several researchers have attempted to predict crowdfunding success using various data-mining techniques. In a study conducted by Greenberg et al. [3], they found that the decision tree classifier predicted the crowdfunding success with an accuracy of 68\% at best, 14\% higher than the related baseline. Mitra and Gilbert [11] predicted the success of crowdfunding utilizing the words and phrases used by the project creators. They analyzed the text data using tools such as Linguistic Inquiry and Word Count (LIWC) [12, 13] to infer psychological meaning. They find that language used in the description contributes to 59\% variance in crowdfunding success. Instead of using static attributes (i.e., attributes available at the launch of the campaign), Etter, Grossglauser, and Thiran [14] combined both direct features and social features to predict the campaign outcome, and their model achieved a 76\% accuracy. Yuan, Lau, and Xu [15] proposed a semantic text analytic approach to predicting crowdfunding success; they found that topic models mined from topic descriptions are useful for prediction. In addition, they found that an ensemble of weak classifiers - random forest performed better than a single strong classifier - support vector machine.

\section{Data and Extracted Features}

\subsection{Data collection}
\begin{figure}
    \centering
    \includegraphics[width = \linewidth]{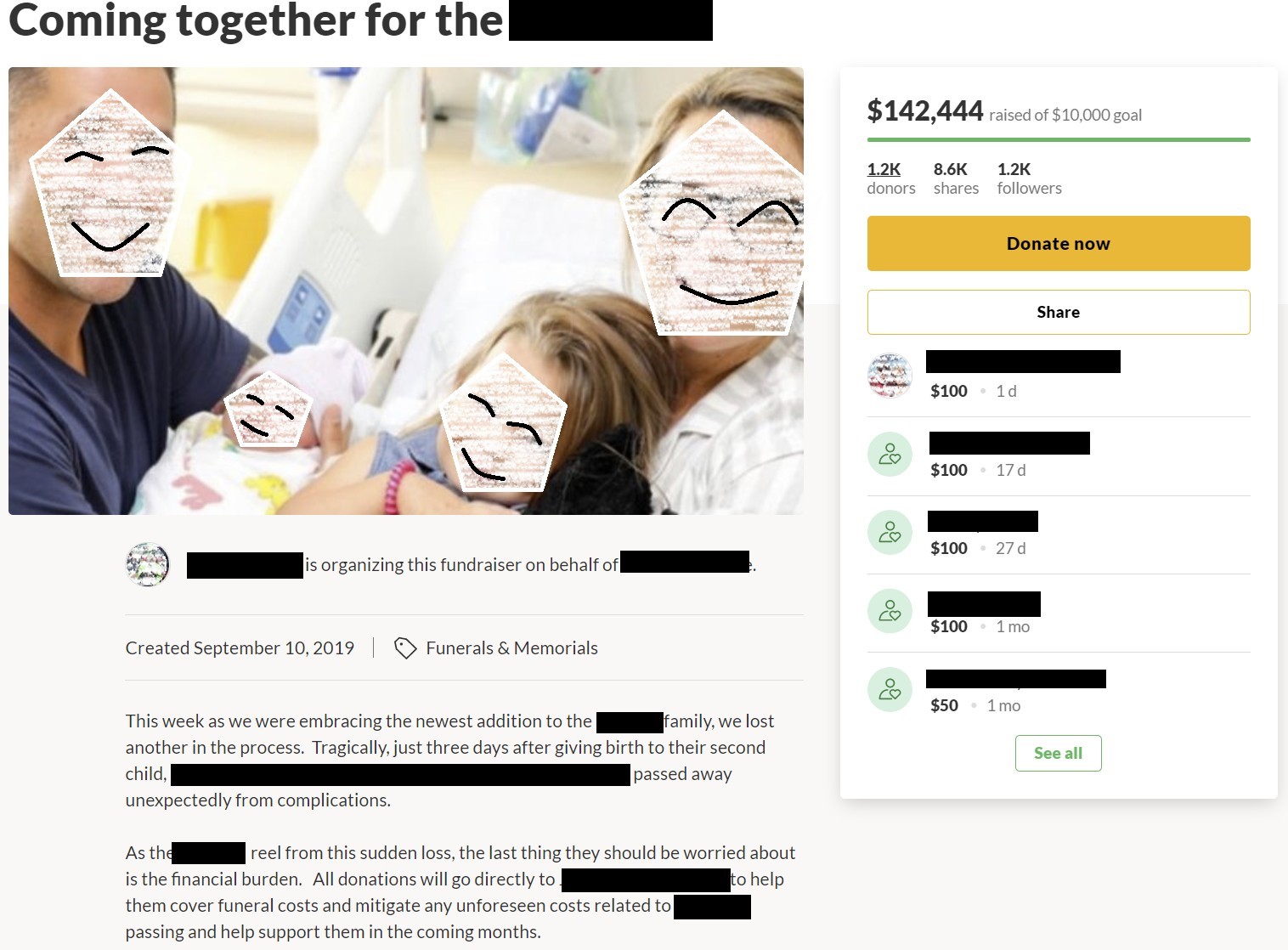}
    \caption{A Campaign on GoFundMe.com (recognizable faces and names are masked to preserve privacy).}
    \label{fig:1}
\end{figure}

\noindent
We first crawl all the available crowdfunding campaigns on GoFundMe. As the example shown in Fig.~\ref{fig:1}, we are able to collect 10,974 crowdfunding campaigns around the world as of November 19, 2019. We decide to use only the U.S. data in our analysis, which amounts to 8,355 U.S. campaigns on GoFundMe.

\subsection{Crawled features}
\noindent
Table~\ref{tab:1} shows the extracted features directly crawled from the website. Dynamic features such as the number of followers, the number of shares, and the number of donors are also included.

\begin{table}[]
    \centering
     \caption{Crawled features.}
     	\vskip 5mm
	\begin{tabular}{c}
		\hline
		Launch Date, Location, Title Cover image, Description,\\  Category, Current Amount, Goal Amount,  \# of Followers,\\ \# of Shares, \# of Donors \\
				\hline
	\end{tabular}
   
    \label{tab:1}
\end{table}

\subsubsection{Inferred Features}
\noindent
Table~\ref{tab:2} shows the inferred features.

\begin{table}[]
    \centering
     \caption{Inferred features.}
     	\vskip 5mm
	\begin{tabular}{c}
		\hline
		The population of the Fundraiser's Location;\\ \# of People in the Cover Image; \\People's Facial Attributes on the Cover Image; \\ Technical and Aesthetic Scores of the Cover Images\\
				\hline
	\end{tabular}
   
    \label{tab:2}
\end{table}

\begin{itemize}
    \item Population: Since the available attributes do not directly list the population, we infer that from the fundraiser’s location (e.g., Los Angeles) using the US Census Bureau data (2018).
    \item Image Quality Assessment: We use a pre-trained model called Neural Image Assessment (NIMA) [17] to predict the aesthetic and technical quality scores for each cover image. The models are trained via transfer learning, where ImageNet pre-trained Convolutional Neural Networks (CNN) are used and fine-tuned for the classification task.
    \item We use Face++\footnote{faceplusplus.com}, which is a face recognition platform based on deep learning. As can be seen from Table~\ref{tab:3}, for each cover image, Face++ returns the following values:
    
    \textbf{Number of faces in the cover image:} the number of faces in the cover image
    
    \textbf{Gender:} male or female
    
    \textbf{Beauty:} attractiveness score given by male and female evaluators, individually
    
    \textbf{Smile:} yes or no
    
    \textbf{Emotion:} contains the values for anger, disgust, fear, happiness, neutral, sadness, surprise
    
    \textbf{Age:} a value between [0,100]
    
    \textbf{Is child:} yes or no, a variable shows whether a child’s face is in the cover image. If a person’s face is detected and the age is under 10

\end{itemize}

\begin{table}[h]
	\centering
	\caption{Examples of the face attributes.}\label{tab:3}
	\vskip 5mm
	\begin{tabular}{|p{0.9in}|p{1.1in}|p{0.8in}|}
		\hline
		&\includegraphics[width=1.1in,height=0.75in]{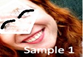}&\includegraphics[width=0.8in,height=0.75in]{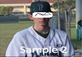}\\
		\hline Gender & Female & Male \\
		\hline Age & 22 & 18 \\
		\hline Emotion (highest score) & Happiness (99.99) & Neutral (75.34) \\
		\hline Beauty & Female: $59.21$ Male: $59.03$ & Female: $75.22$ Male: $73.9$\\
		\hline Smile & Yes & No \\
		\hline
	\end{tabular}
\end{table}

\section{Methodology}
\subsection{Crowdfunding success metrics}
\begin{figure}
    \centering
    \includegraphics[width = \linewidth]{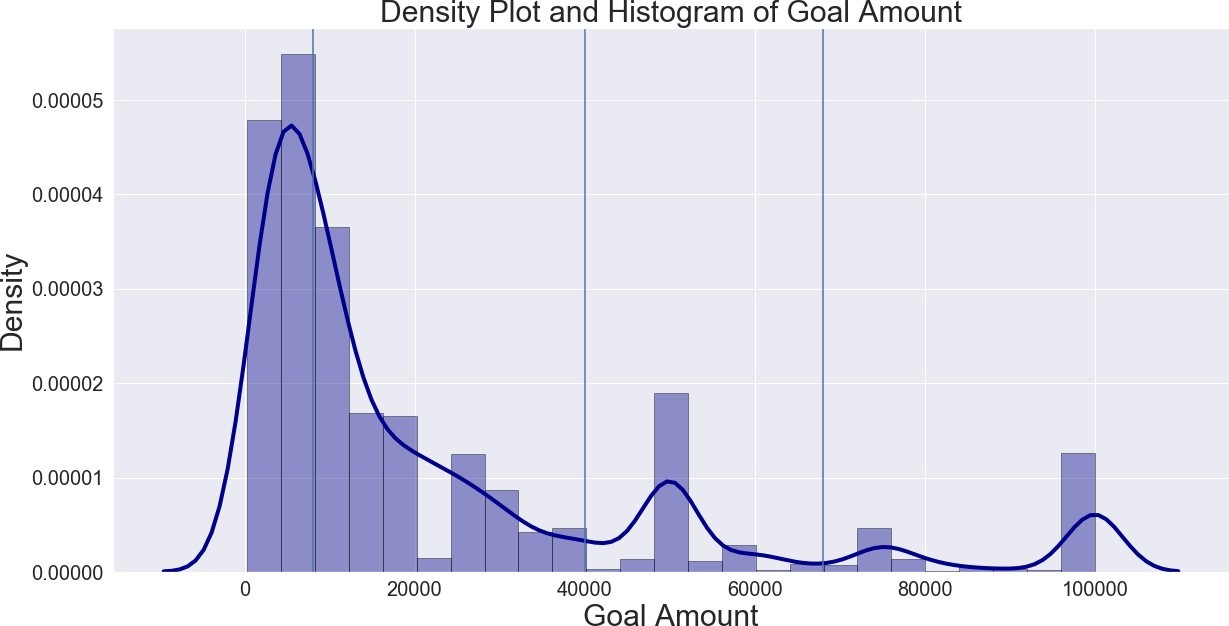}
    \caption{Histogram of the goal amount.}
    \label{fig:2}
\end{figure}

\begin{figure}
    \centering
    \includegraphics[width = 0.97\linewidth]{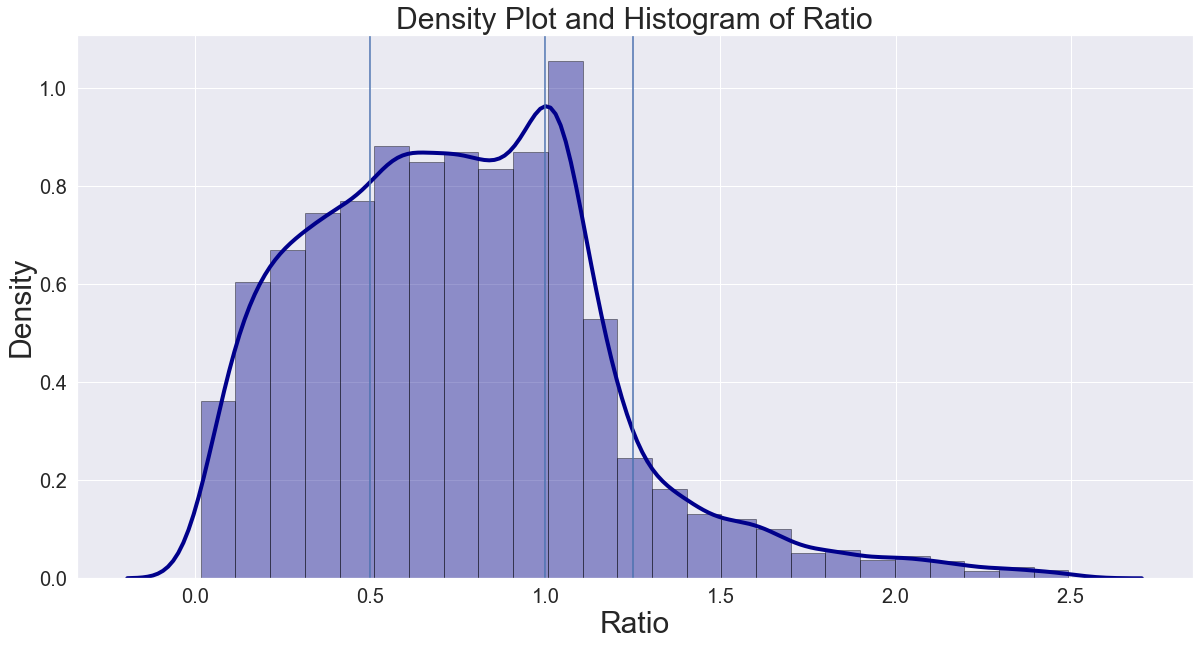}
    \caption{Histogram of the ratio.}
    \label{fig:3}
\end{figure}

\noindent
We bin the data into four groups according to the smoothed goal amount shown in the distribution in Fig.~\ref{fig:2}, which reveals four distinctive groups: (0, 8000], (8000, 40000], (40000, 68000], and (68000, 100000]. In each group, we define the success of a crowdfunding campaign using the ratio of the amount of money that has been raised so far to the fundraiser’s goal amount. Similar to setting the thresholds for the goal amounts, the ratio (Fig.~\ref{fig:3}) is empirically binned into four groups: (0, 0.5], (0.5, 1], (1, 1.25], and (1.25, 2.5]. Most of the ratios of campaigns are lower than 2.5 (4.19\%). To have a better understanding of the generalized effects that contribute to a crowdfunding campaign’s success, we drop the fundraisers with ratios greater than 2.5. Fundraisers with ratios from 0 to 0.5 are defined as “highly unsuccessful (-2)”; ratios from 0.5 to 1 are defined as “unsuccessful (-1)”; while ratios from 1 to 1.25 are defined as “successful (+1)”; ratios from 1.25 to 2.5 are defined as “highly successful (+2)”.

\begin{equation*}
    Ratio = \frac{Money\:That\:Has\:Been\:Raised\:So\:Far}{Goal\:Amount}
\end{equation*}

\subsection{Image features}
\begin{figure}
    \centering
    \includegraphics[width = \linewidth]{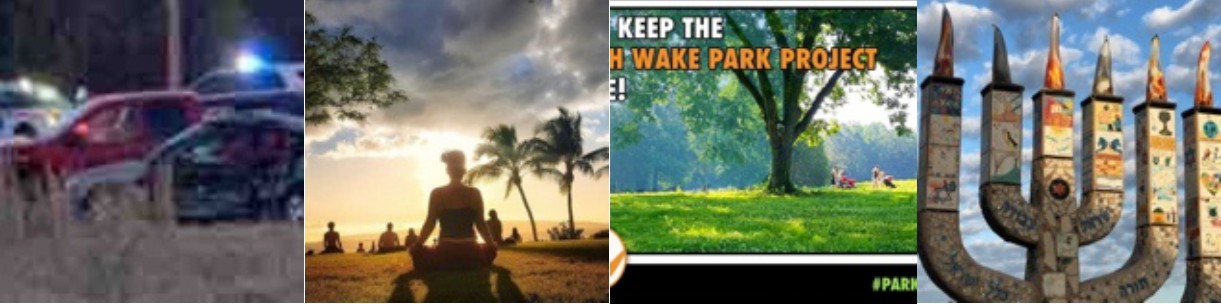}
    \caption{Examples of aesthetic score prediction by the MobileNet (the score goes up from left to right).}
    \label{fig:4}
\end{figure}

\begin{figure}
    \centering
    \includegraphics[width = \linewidth]{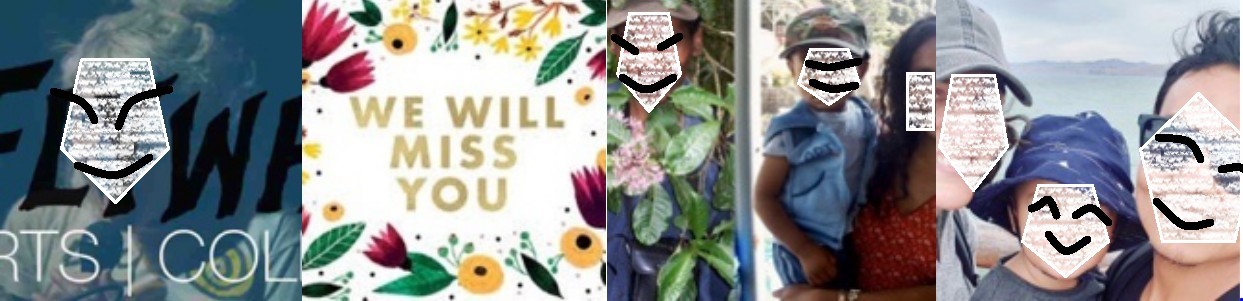}
    \caption{Examples of technical score prediction by the MobileNet (the score goes up from left to right).}
    \label{fig:5}
\end{figure}

\noindent
We use a pre-trained model called NIMA [17] to predict the aesthetic quality and technical quality of cover images, respectively. The models are trained via transfer learning, where ImageNet pre-trained CNNs are used and fine-tuned for the image quality classification task. As shown in Fig.~\ref{fig:4}, the predictions indicate that the aesthetic classifier correctly ranks the cover images from very aesthetic (the rightmost art image) to the least aesthetic (the leftmost image with two cars). The higher the aesthetic score is, the more aesthetic the image is. Similarly, Fig.~\ref{fig:5} shows that the technical quality classifier predicts higher scores for visually pleasing images (third and fourth from the left) versus the images with JPEG compression artifacts (second) or blur (first).

\begin{figure}
    \centering
    \includegraphics[width = \linewidth]{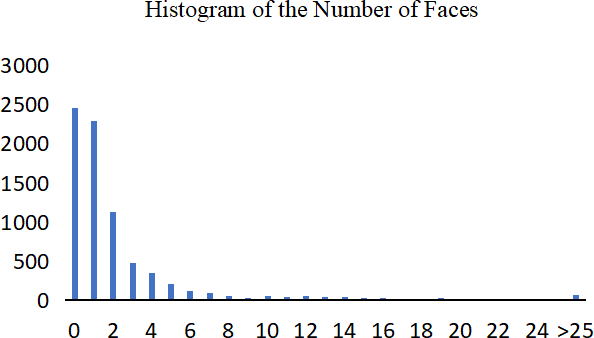}
    \caption{Face attributes.}
    \label{fig:6}
\end{figure}

In order to better understand the influence of facial attributes on crowdfunding success, we use the prediction outcomes from Face++. Fig.~\ref{fig:6} shows the summary statistics of the Face++ results.

\subsection{Text features}
\noindent
Similar to Wu et al. [18] and Zhang et al. [19], we compute 92 LIWC features (e.g., word categories such as “social” and “affect”) to model the text Data [13]. These features can potentially reflect the distribution of the text data.

\subsection{Fusion methods}
\begin{figure}
    \centering
    \includegraphics[width = \linewidth]{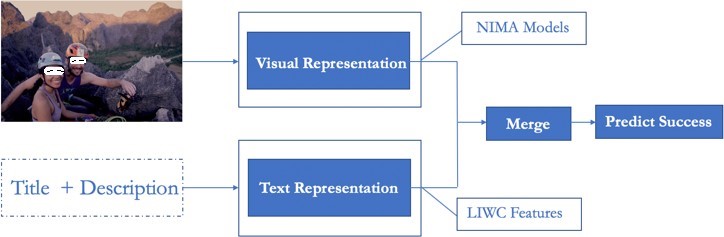}
    \caption{Flowchart of the data fusion.}
    \label{fig:7}
\end{figure}

\noindent
Based on the ratio of the amount of money that has been raised so far and the fundraiser’s goal amount, campaigns are separated into four classes – “highly unsuccessful (- 2)” ((0, 0.5]), “unsuccessful (-1)” ((0.5, 1]), “successful (+1)” ((1, 1.25]), “highly successful (+2)” ((1.25, 2.5]). We apply both early fusion and late fusion [16] to predict crowdfunding success using pictorial and textual features. Fig.~\ref{fig:7} shows the flowchart of the multimodal data fusion. Title and description are used as text representation, and the profile images are employed as a visual representation. We use the NIMA models to get image feature scores and the LIWC models to get text feature scores. We predict the success of crowdfunding campaigns by merging image and text features. In recent years, using ensembles of classifiers has attracted a lot of attention in the research community [20, 21]. For instance, Zhu, Yeh, and Cheng [22] show that the fusion classifier of image and text has higher accuracy than the state-of-art methods for image classification.

\section{Experiments and Discussions}
\noindent
In this section, we investigate the relationship between the fundraiser’s success and the attributes of the fundraiser. First, we analyze the category proportion of each goal amount group. After controlling for the category, we dig deep into the city population, the LIWC features, and the image quality.

\subsection{Campaign category}
\noindent
Each campaign on GoFundMe belongs to one of nineteen unique categories (e.g., Weddings \& Honeymoons). The number of campaigns in each category is evenly distributed with the exception of “Other” and “Non-Profits \& Charities.” However, the proportions of successful and unsuccessful campaigns in each category are not uniform.
 
“Weddings \& Honeymoons,” “Competitions \& Pageants,” and “Travel \& Adventure” are the top three categories that are most likely to fail in a fundraising campaign that has a goal amount between \$0 and \$8000. The top three categories that are most likely to succeed in a fundraiser from that goal amount group are “Volunteer \& Service,” “Dreams, Hopes \& Wishes,” and “Celebrations \& Events.” In the goal amount between \$8000 and \$40000 group, the top three “unsuccessful” categories become “Business \& Entrepreneurs,” “Missions, Faith \& Church,” and “Sports, Teams \& Clubs.” The top three “successful” categories are “Babies, Kids \& Family,” “Accidents \& Emergencies,” and “Funerals \& Memorials.” Regardless of the goal amount, the categories that are most likely to succeed are health-related. As the goal amount increases, “Medical, Illness \& Healing” and “Funerals \& Memorials” remain the two categories with the highest likelihood of success. More donations are made to the events related to health.

\subsection{Population}
\noindent
We analyze the fundraiser’s city population for each category. We compare the mean ratios of the campaigns in cities with small populations and the campaigns in cities with large populations. After performing the Student’s t-test, We find that only “Babies, Kids \& Family” is influenced by the population of the fundraiser’s city population, and it is only significant in the \$8000 to \$40000 goal amount group ($t = 3.25, p < 0.05$). The mean success ratio in small towns is significantly higher ($t = 2.25, p < 0.05$) than that in big cities for this category. This suggests that it is easier for fundraisers to achieve their fundraising goals if they are from a smaller city. We dig deeper by taking a look at the number of times a fundraiser gets shared via social media. The number of shares is also available from the website of GoFundMe. The fundraiser from a small city raising money for “Babies, Kids \& Family” has significantly more shares than the one from a big city ($p < 0.05$). A possible hypothesis is that the people from a small city have a stronger sense of belonging or community than the people from a large city.

\subsection{Campaign description}
\noindent
Since the category is the main variable to analyze, we examine the LIWC features in each category so as to control the influence of the difference between categories. As we expect, some categories of the fundraising campaigns are influenced by the LIWC features.

Table~\ref{tab:4} shows the LIWC features that have a significant influence ($p<5.4E-4$, Bonferroni-corrected) on crowdfunding’s performance. In the (\$0, \$8000] goal amount group, the results of the Pearson correlation indicate that there are significant associations between crowdfunding success and the project’s descriptions. 

After performing the Pearson correlation test, we find that “Animals \& Pets” and “Competitions \& Pageants” are correlated to the way the project description is written ($p < 0.0001$). For the “Animals \& Pets” category, “insight” is negatively correlated with the success of a crowdfunding the campaign, which suggests that people are more likely to donate to animals and pets if the description includes description includes words such as “think,” “know,” or “believe.”

GoFundMe allows people to raise money for someone else. That is why the description is not always written by the people who actually need help and is also why the description is not always written using the first-person pronoun. For the “Competitions \& Pageants” category, “Clout” and “they” are negatively correlated with success. This suggests that if someone wants to raise some money for their competitions or pageants, they should write the description from his/her perspective and try to avoid asking someone else to write the fund description for them. They should also try to write the description in a more humble way.

The reason that “bio” and “health” are positively correlated with the chance of success of “Sports, Teams \& Clubs” is that these words are more related to health issues. This suggests that probably the person behind the fundraiser is likely in need of medical treatment. As we saw before, the fundraisers about medical treatments are always more likely to receive donations.

In the (\$8000, \$40000] goal amount group, even more categories are influenced by the LIWC features. “shehe” and “social” are positively correlated with the success of a fundraiser in the “Community \& Neighbors” category. In addition, a description that seems more anxious or more worried is more likely to help the fundraiser raise more money if it is about “Missions, Faith \& Church”.

“Focusfuture” is positively related to a fundraiser’s success in the “Dreams, Hopes \& Wishes” category. When people write a description of dreams, they should focus more on the future.

A higher “Clout” value indicates a more confident description, and a lower value indicates a more humble description. As the proposed analysis shows, categories like weddings and honeymoons are not easy to get the donation. People probably favor humble couples. 

For the “Sports, Teams \& Clubs” category, it is counter-intuitive that the descriptions with more “achieve” words have a negative influence on donation. Normally, people would love to see someone with the ambition and desire to win when it comes to sports. It is really interesting that based on our data, those people actually receive less donation. For this part, we still do not have a convincing explanation other than the suspicion that overstating may actually turn third-party people off.

We have not found enough evidence to conclude that there is a relationship between the description and the success of a fundraiser in the (\$40000, \$68000] or the (\$68000, \$100000] group.

\begin{table*}[h]
	\centering
	\caption{LIWC features that have significant influences.}\label{tab:4}
		\vskip 5mm
	\begin{tabular}{|c|c|c|c|c|c|c|c|}
		\hline
		\textbf{Goal }                         & \textbf{Category }                                 &\textbf{LIWC}         & \textbf{Examples  }                  & \textbf{Mean}  & \textbf{SD}    & \textbf{r}      & \textbf{p-value}  \\
		\hline
		\multirow{4}{*}{\$0-\$8000}     & \multirow{2}{*}{Competitions \& Pageants} & Clout        & High: confident Low: humble & 73.02 & 21.84 & -0.181 & 0.0002   \\
		\cline{3-8}
		&                                           & social       & mate, talk, they            & 9.96  & 4.02  & -0.168 & 0.0005   \\
			\cline{2-8}
		& Animals \& Pets                           & insight      & think, know                 & 1.51  & 0.87  & -0.304 & 2.80E-05 \\
			\cline{2-8}
		& Sports, Teams \& Clubs                    & bio          & eat, blood, pain            & 0.86  & 0.95  & 0.275  & 0.0002   \\
			\cline{1-8}
		\multirow{6}{*}{\$8000-\$40000} & \multirow{2}{*}{Community \& Neighbors}   & shehe        & she, her, him               & 1.46  & 2.23  & 0.220   & 4.70E-05 \\
			\cline{3-8}
		&                                           & Social       & mate, talk, they            & 12.75 & 4.68  & 0.200    & 0.0002   \\
			\cline{2-8}
		& Dreams, Hopes \& Wishes                   & focus future & may, will, soon             & 1.39  & 0.95  & 0.256  & 0.0001   \\
			\cline{2-8}
		& Missions, Faith \& Church                 & anx          & worried, fearful            & 0.11  & 0.24  & 0.305  & 1.20E-06 \\
			\cline{2-8}
		& Weddings \& Honeymoons                    & Clout        & High: confident Low: humble & 93.63 & 10.50  & -0.390  & 0.0004   \\
			\cline{2-8}
		& Sports, Teams \& Clubs                    & achieve      & win, success, better        & 4.26  & 2.34  & -0.196 & 0.0003   \\
					\cline{1-8}

	\end{tabular}
\end{table*}

\subsection{Image quality}
\begin{figure}
    \centering
    \includegraphics[width = \linewidth]{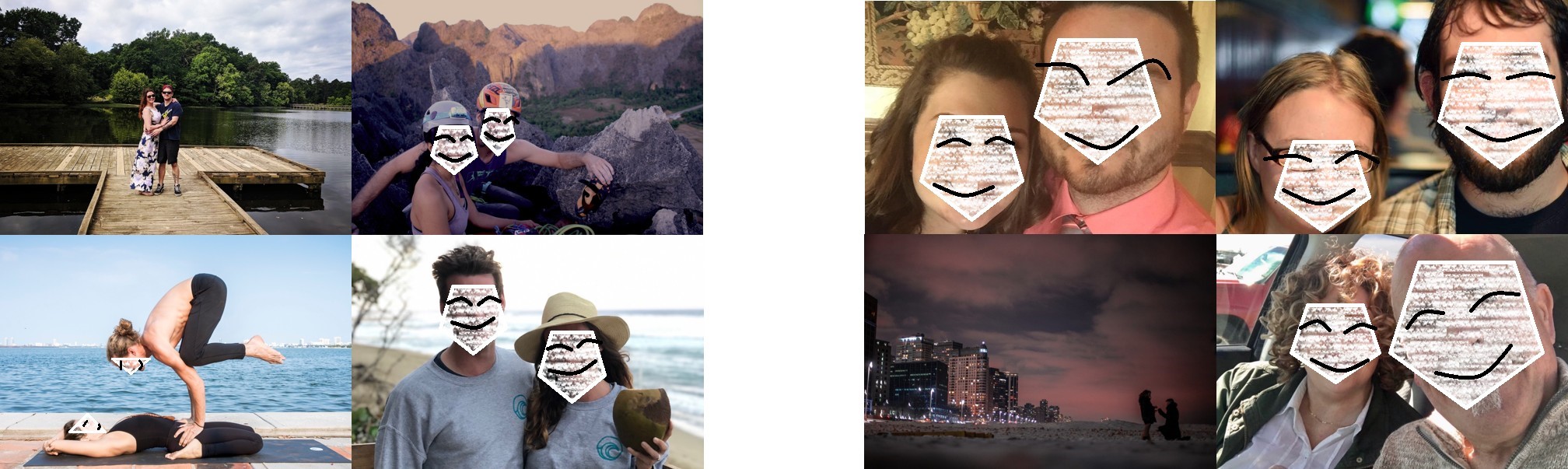}
    \caption{Randomly selected photos from the successful (left) and unsuccessful (right) groups.}
    \label{fig:8}
\end{figure}

\noindent
Table~\ref{tab:5} shows the image quality features that have a significant influence ($p<0.025$, Bonferroni-corrected) on the fundraiser’s performance. We find that “Competitions \& Pageants,” “Community \& Neighbors,” and “Weddings \& Honeymoons” are the only three categories that are influenced by the image quality. This suggests that the success of fundraisers is related to their cover image quality if their fundraising purpose falls into one of these three categories. Take the “Weddings \& Honeymoon” category as an example. Fig.~\ref{fig:8} shows several randomly selected photos from the “highly successful (+2)” and “highly unsuccessful (-2)” groups. The ones from the “highly successful (+2)” group are on the left, and the ones from the “highly unsuccessful (-2)” group are on the right. By looking at them, we find that campaigns with higher success rates have more aesthetic cover images and the fundraisers seemed to make conscious efforts in taking those photos. In contrast, campaigns with lower success rates have casual selfie cover images. However, it is still unclear why the image quality is negatively correlated to the success of fundraisers in the “Competitions \& Pageants” category. One hypothesis is that overdoing the cover photos makes people think that the activity should already be well funded. Future work could investigate whether there is a non-linear relationship between image quality and campaign success by combining the images of all categories.

\begin{table*}[h]
	\centering
	\caption{Image quality features that have significant influences.}\label{tab:5}
	\vskip 5mm
	\begin{tabular}{|c|c|c|c|c|c|c|}
		\hline
		\textbf{Goal}                          & \textbf{Category}                                  & \textbf{Image Info }                           & \textbf{Mean}  & \textbf{SD}    & \textbf{r}      & \textbf{p-value}  \\
		\hline
		\multirow{4}{*}{\$8000-\$40000}     & \multirow{2}{*}{Competitions \& Pageants} & aesthetic score        & 5.00 & 0.40  & -0.355 & 0.0002 \\
		\cline{3-7}
		& &technical score & 5.72 & 0.53 & -0.295 & 0.0023 \\
		\cline{2-7}
		& Community \& Neighbors & technical score &	5.28&	0.78	&0.174&	0.0013\\
		\cline{2-7}
		&Weddings \& Honeymoons	&aesthetic score&	4.65&	0.56&	0.320&	0.0042\\
		\hline
	\end{tabular}
\end{table*}

\subsection{Face attributes}
\noindent
We perform the Pearson correlation test. Table~\ref{tab:6} shows face attribute features that have significant influences ($p<0.02$, Bonferroni-corrected) on the crowdfunding’s performance in some categories. In the (\$0, \$8000] goal amount group, a smaller number of faces corresponds to a higher chance to succeed in crowdfunding related to competitions and pageants. In fact, the mean number of faces of the “highly unsuccessful (-2)” group where the ratio is between 0 and 0.5 is 4.07, and that of the “highly successful (+2)” group where the ratio is between 1.25 and 2.5 is 2.41. In the (\$8000, \$40000] goal amount group, we find that the number of faces and age are positively correlated with crowdfunding performance if it is about medical, illness, and healing. In the “highly successful (+2)” group, the mean number of faces and the age is 1.81 and 34.31, respectively. For the “highly unsuccessful (-2)” group, the mean is 1.00 and 21.50, respectively. We analyze some cover images and find that donors respond positively to the family photos which were taken before the accidents happened. However, what is interesting here is that the number of faces does not work the same way in the competitions and pageants category as it does in the medical and healing category. We think the reason is that they are different categories. Recall in the previous sections; we find that medical fundraisers are almost always the most successful. People do care about others and are willing to donate if it is urgent or about life and death. People are less likely to donate to less urgent events like weddings and competitions. For “Animals \& Pets” and “Travel \& Adventure,” we find that the appearance of a child boosts donations.

\begin{table*}[h]
	\centering
	\caption{Facial features that have significant influences.}\label{tab:6}
	\vskip 5mm
	\begin{tabular}{|c|c|c|c|c|c|c|}
		\hline
		\textbf{Goal}                          & \textbf{Category}                                  & \textbf{Face++}                            & \textbf{Mean}  & \textbf{SD}    & \textbf{r}      & \textbf{p-value}  \\
		\hline
		\$0-\$8000& 	Competitions \& Pageants&	Num face&	3.39&	5.48&	-0.141         &             	0.0037 \\
		\hline
		\multirow{5}{*}{\$8000-\$40000}     & \multirow{2}{*}{Medical, Illness \& Healing} & Num face       & 1.36 & 1.31  & 0.344 & 0.0082 \\
		\cline{3-7}
		& &Age  & 27.17 & 10.39 & 0.349 & 0.0073 \\
		\cline{2-7}
		& Animals \& Pets &is child &	0.01&	0.09	&0.474&	2.3E-16\\
		\cline{2-7}
			& Travel \& Adventure &is child &	0.02&	0.15	&0.458&	1.2E-05\\
		\cline{2-7}
	&	Missions, Faith \& Church&is child &	24.18&	18.97	&0.168&	0.0085\\
		\hline
		
	\end{tabular}
\end{table*}

\subsection{Classification evaluation}
\noindent
In this section, we conduct four-class classification experiments to evaluate the effectiveness of our proposed features using the Random Forest model. We separate the data into a training set and a testing set. 90\% of data are used as the training data, and 10\% are used as the testing data. The target we try to predict is the range of the ratio of a fundraiser. The outcome is “highly unsuccessful (-2)” ((0, 1.25]), “highly successful (+2)” ((1.25, 2.5]). Based on the aforementioned analysis, we choose features with significant correlation as the input to perform the classification for each goal amount group. In the (\$0, \$8000] goal amount group, we find the LIWC features and Face++ features are correlated with the ratio. In the (\$8000, \$40000] goal amount group, LIWC features, Face++ features, the population, and the image quality are found to have an impact on the ratio. In the (\$40000, \$68000] and (\$68000, \$100000] goal amount groups, we do not find significant differences in the features among different ratio groups.

\begin{itemize}
    \item \textbf{Basic Information with Category.} The input of the model only includes basic information like launch date, city, state, and category information. This is the baseline model.
    \item \textbf{Single LIWC.} The input of the model only includes LIWC features of each category. Specifically, we choose different LIWC features in each category based on our proposed analysis.
    \item \textbf{Single City Population.} The input of the model only includes the city population of each category. Specifically, we choose the data of category based on our proposed analysis.
    \item \textbf{Single Face++.} The input of the model only includes the Face++ features of each category. Specifically, we choose different Face++ features in each category based on our proposed analysis.
    \item \textbf{Single Image Quality.} The input of the model only includes the image quality features of each category. Specifically, we choose different image quality features in each category based on our proposed analysis.
    \item \textbf{Early Fusion.} The text description and image information are combined as the input.
    \item \textbf{Late Fusion.} The text description and image information are used to construct separate models, respectively, before a final decision is combined.
\end{itemize}

For all the above settings, we employ Random Forest as the classifier due to its empirically good performance. We show the quantitative results of our experiments in Table~\ref{tab:7}. The performance of different models is evaluated by four metrics: accuracy, precision, recall, and F1-measure. During the experiments, the number of estimators is set to 1000, and the minimum number of samples required to split an internal node is set to 2 for every setting. Each set is conducted using 10-fold cross-validation. We conduct experiments in each goal amount group and calculate the weighted metrics.

\begin{table*}[h]
	\centering
	\caption{Comparison of accuracy, precision, recall, and F1-score.}\label{tab:7}
	\vskip 5mm
	\begin{tabular}{|c|c|c|c|c|c|}
		\hline
		\textbf{Goal Amount}                          & \textbf{}                                  & \textbf{Accuracy}                            & \textbf{Precision}  & \textbf{Recall}    & \textbf{F1-score}        \\
		\hline
		\multirow{5}{*}{\$0-\$8000}     & Basic &	0.40&	0.38&	0.40	&0.38 \\
		\cline{2-6}
		&LIWC&	0.45&	\textbf{0.41} &	0.45&	0.41 \\
			\cline{2-6}
		&Face++	&0.44&	0.34&	0.44&	0.37 \\
			\cline{2-6}
		&Basic+LIWC+Face++&	\textbf{0.46}&	0.40&	\textbf{0.46}&	\textbf{0.41}\\
			\cline{2-6}
		&Late Fusion	&0.43&	0.38&	0.43&	0.39\\
			\hline
		\multirow{6}{*}{\$8000-\$40000}     & Basic &	0.49&	0.46&	0.49	&0.47 \\
		\cline{2-6}	
		& LIWC	&0.47&	0.42&	0.47&	0.44\\
			\cline{2-6}	
		& Population	&0.48&	0.23&	0.48&	0.31 \\
			\cline{2-6}	
	&	Image Quality	&\textbf{0.58}	&\textbf{0.55} &	\textbf{0.58}	&\textbf{0.56} \\	
				\cline{2-6}	
	&	B+LIWC+F+P+I&	0.55 &	0.42 &	0.55&	0.46\\
		\cline{2-6}	
	&Late Fusion&	0.53	&0.43&	0.53&	0.46 \\
	\hline
		\$40000-\$68000 &	Basic&	0.72&	0.67&	0.72&	0.69\\
		\hline
		\$68000-\$100000 &	Basic	&0.61	&0.57&	0.61&	0.58\\
		\hline
		\multirow{3}{*}{Total (Weighted)}   &  Basic	&0.49&	\textbf{0.46}&	0.49	&\textbf{0.46}  \\
			\cline{2-6}	
			&Early Fusion&	\textbf{0.51}&	0.41	&\textbf{0.51}&	0.44 \\
				\cline{2-6}	
				&Late Fusion&	0.48&	0.41&	0.48&	0.43\\
	
		\hline
		
	\end{tabular}
\end{table*}

As Table~\ref{tab:7} shows, the baseline models with the basic information are the worst at prediction in each group. In contrast, adding extra information can always increase classification performance.

In the (\$0, \$8000] group, the performance of the models using the basic information and the models using LIWC or Face++ is quite comparable. The one that combines all of the features is the best according to most metrics.

In the (\$8000, \$40000] group, the city population is not really useful during the classification. The aesthetic score and technical score of the cover image are really helpful for making a classification. The result is consistent with the study of Cheng et al. [24], which shows that image features significantly improve success prediction performance, particularly for crowdfunding projects with a little text description. It is a surprise that the Single Image Quality setting is the best at classification.

Since there is no sufficient evidence to conclude the relationship between features and success within the (\$40000, \$68000] and (\$68000, \$100000] groups, we use the basic information as the input. As we can see from Table~\ref{tab:7}, it is easier to make a reliable classification within a higher goal amount group.

We intend to analyze the factors of the fundraiser campaign’s success in two steps. First, we screen the potential factors by conducting multiple correlation analyses. Second, we use the factors we choose from the first step to build the classifiers and evaluate their effects on the basis of whether they improve or worsen the classification performance. One alternative approach could be building the classifier directly and letting the classifier choose variables automatically.

\section{Conclusions}
\noindent
In this study, we focus on understanding and predicting the performance of the crowdfunding campaigns on GoFundMe, which is diverse in funding categories and charity-minded. We analyze the attributes available at the launch of the campaign and identify attributes that are important for the major campaign categories. Furthermore, we have bench-marked several computational models and identified a multimodal fusion classifier that significantly improves the prediction result. We believe that the findings and models from this study provide effective mechanisms to make a crowdfunding campaign successful in different categories.

\renewcommand\refname{\zihao{5}

\begin{strip}
\end{strip}

\begin{strip}
\end{strip}

\mbox{}
\clearpage
\clearpage
\zihao{5}
\noindent

  \end{document}